\documentclass[%
 reprint,
 amsmath,amssymb,
 aps,
]{revtex4-1}

\usepackage{graphicx}
\usepackage{bm}% bold math

\begin{document}

\title{Scalable Neutral Atom Quantum Computer
with Interaction on Demand}

\author{Mikio Nakahara}\email{nakahara@math.kindai.ac.jp}
\affiliation{
Research Center for Quantum Computing, Interdisciplinary Graduate School of Science and Engineering, Kinki University, 
3-4-1 Kowakae, Higashi-Osaka, 577-8502, Japan}
\affiliation{
Department of Physics, Kinki University, 
3-4-1 Kowakae, Higashi-Osaka, 577-8502, Japan}
\author{Tetsuo Ohmi}\email[]{ohmi@math.kindai.ac.jp}
\affiliation{
Research Center for Quantum Computing, Interdisciplinary Graduate School of Science and Engineering, Kinki University, 
3-4-1 Kowakae, Higashi-Osaka, 577-8502, Japan}
\author{Yasushi Kondo}\email[]{kondo@phys.kindai.ac.jp}
\affiliation{
Research Center for Quantum Computing, Interdisciplinary Graduate School of Science and Engineering, Kinki University, 
3-4-1 Kowakae, Higashi-Osaka, 577-8502, Japan}
\affiliation{
Department of Physics, Kinki University, 
3-4-1 Kowakae,\, Higashi-Osaka, 577-8502, Japan}
%\thanks{}
%\affiliation{}


\date{\today}

\begin{abstract}
We propose a scalable neutral atom quantum computer with an on-demand
interaction. Artificial lattice of near field optical traps
is employed to trap atom qubits. Interactions between atoms
can be turned off if the atoms are separated by a high enough
potential barrier so that the size of the atomic wave function
is much less than the interatomic distance. 
One-qubit gate operation is implemented by a gate control laser beam 
which is attached to an individual atom. 
Two-qubit gate operation between a particular pair of atoms 
is introduced by leaving these atoms in an optical lattice
and making them collide so that a particular two-qubit state
acquires a dynamical phase. Our proposal is feasible within existing 
technology developed in cold atom gas, MEMS, nanolithography, and various 
areas in optics. 
\end{abstract}

%\keywords{}

\maketitle

\section{Introduction}

Quantum computing is an emerging discipline in which information
is stored and processed by employing a quantum system. In spite of 
many proposals for potentially scalable quantum computers \cite{roadmap,
nakaharaohmi}, most
physical realizations so far accommodate qubits on the order of ten, at most.
One of the obstacles against scalability in the previous
proposals is the absence of controllable interaction between
an arbitrary pair of qubits. NMR realization of a quantum computer
employs nuclei in a molecule as qubits,
whose inter-qubit coupling is difficult to control \cite{nakaharaohmi}
(see, however, \cite{yasuda} for a controllable spin-spin coupling
and \cite{esr} for a fullerene-based ESR quantum computer). 
Many pulses had to be applied
to eliminate unwanted interactions in the
demonstration of the Shor's algorithm with an NMR
quantum computer \cite{nmrshor}. 

It is the purpose of the present paper to propose a new design of a
neutral atom quantum computer with an on-demand interaction,
which is potentially scalable up to a large number of qubits
within currently available technology. 
Neutral atoms are believed to be robust against decoherence, another
obstacle against a physical realization of a working quantum computer, since
they are electrically neutral. 
%Each qubit in our proposal has its own trap and 
%gate control laser beams.
% and two-qubit interaction
%between an arbitrary pair of qubits can be introduced on demand.  
Each atom is trapped by a near field Fresnel diffraction
light, abbreviated as
NFFD light hereafter, which 
is produced by letting the trap laser light pass through an aperture with
a diameter comparable to the wavelength of the laser.
Apertures are arranged in a two-dimensional regular 
lattice, for example. Nevertheless, the
arrangement can be designed freely as will be shown later.
Each lattice site is equipped with its 
own gate control laser beam so that individual single-qubit gate operations 
can be applied on many qubits individually and
simultaneously. A controlled two-qubit
gate operation between an arbitrary pair of qubits is made possible by
selectively
leaving two atoms in a one-dimensional optical lattice
and making particular two-qubit states collide by conveying them with
a pair of hyperfine state-dependent optical lattice potentials. 
The corresponding two-qubit subspace acquires a dynamical phase, which
is controllable by adjusting the collision time.
The qubits are sent back to their initial positions after
entanglement is established by this state-selective collision \cite{mandel2}. 

Technology required for physical implementation of the proposed 
quantum computer
has already been developed in the areas of cold atom trap, near field
optical scanning microscope (SNOM), nanolithography,
and quantum optics, among others, 
and we believe there should be no obstacles against its physical realization.

We outline the design of a neutral atom quantum computer proposed by
Mandel {\it et al.} in the next section. We point out problems
inherent in their design. Individual atom trap by
making use of NFFD light is introduced in Section III. 
Section IV is devoted to 1- and 2-qubit gate implementations.
We propose alternative designs of a neutral atom quantum computer,
which works under the same principle in Section V. Summary
and discussions are given in Section VI.

\section{Neutral Atom Quantum Computer in Optical Lattice}

Suppose an atom is put in a laser beam with an oscillating electric field
$\bm{E}(\bm{x}, t) = {\rm Re} (\bm{E}_0(\bm{x}) e^{-i \omega_L t})$,
where $\omega_L$ is the laser frequency. It is assumed that $\omega_L$ is
close to some transition frequency $\omega_0 =E_e-E_g$ 
between two states $|g \rangle$
and $|e \rangle$ of the atom. The interaction between the electric field
and the dipole moment of the atom introduces an interaction Hamiltonian
\begin{equation}
H_i = -\frac{1}{2} (\bm{E}_0 \cdot \bm{d}) (e^{-i \omega_L t}
+ e^{i \omega_L t}),
\end{equation}
where $\bm{d}$ is the dipole moment operator of the atom.
This interaction introduces an effective potential of the form
\begin{equation}
V(\bm{x}) = \frac{\hbar |\Omega_{eg}(\bm{x})|^2}
{4\Delta_{eg}}
\end{equation}
for an atom in the ground state, which is called the AC Stark shift.
Here $\Omega_{eg}=\langle e|\bm{d}|g\rangle \cdot \bm{E}_0(\bm{x})/\hbar$, while
$\Delta_{eg} = \omega_L-\omega_0$ is the detuning. 
We have ignored the small natural line width of the excited state. 
In case $\Delta_{eg}>0$
(blue-detuned laser), $V(\bm{x})$ is positive and a region with large
$V(\bm{x})$ works as a repulsive potential. In contrast, if $\Delta_{eg}<0$
(red-detuned laser), $V(\bm{x})$ is negative and a region with large
$|V(\bm{x})|$ works as an attractive potential. We use these facts extensively
in the following proposals.

It is possible to confine neutral atoms 
by introducing a set of counter propagating laser beams with the same
frequency and amplitude along the $x$-axis. 
These beams produce a standing wave potential of the form
\begin{equation}
V_{\rm ol}(x) = V_{0x} \cos^2 (kx),
\end{equation}
where $k$ is the wave number of the laser. It is possible to confine
atoms three-dimensionally by adding two sets of counterpropagating
beams along the $y$- and the $z$-axes, which results in
\begin{equation}
V_{\rm ol}(\bm{x}) 
= V_{0x} \cos^2(kx) + V_{0y} \cos^2 (ky) + V_{0z} \cos^2 (kz), 
\end{equation}
where we put the wave numbers of the six lasers are identical for
simplicity. Note that the lattice constant is always $\lambda/2$,
where $\lambda$ is the wavelength of the laser.

Mandel {\it et al.} trapped atoms in such an optical potential.\cite{mandel2}
They used the Rabi oscillation to implement one-qubit gates, in which
a microwave (MW) field is applied. Naturally, all the atoms are
under MW irradiation and selective one-qubit gate operation is impossible.
They have chosen qubit basis vectors 
$|0 \rangle =|F=1, m_F=-1 \rangle$ and $|1 \rangle =|F=2, m_F=-2 \rangle$
to be in harmony with the one-qubit gate operation making use of the
Rabi oscillation. 

For two-qubit gate operations, they introduced time-dependent
polarization in the counterpropagating laser beams as
\begin{equation}
\begin{array}{c}
\bm{E}_+(\bm{x}) = e^{ikx}(\hat{\bm{z}} \cos \theta +
\hat{\bm{y}} \sin \theta),
\vspace{.2cm}\\
\bm{E}_-(\bm{x}) = e^{-ikx}(\hat{\bm{z}} \cos \theta -
\hat{\bm{y}} \sin \theta).
\end{array}
\end{equation}
These counterpropagating laser beams produce an optical potential
of the form
\begin{equation}
\bm{E}_+(\bm{x})+\bm{E}_-(\bm{x}) \propto \sigma^+ \cos (kx-\theta)
+ \sigma^- \cos (kx+\theta),
\end{equation}
where $\sigma^+$ ($\sigma^-$) denotes counterclockwise (clockwise)
circular polarization. Note that the first component ($\propto \sigma^+$)
moves along the $x$-axis as $\theta$ is increased, while the
second component ($\propto \sigma^-$) moves along the $-x$-axis under
this change. \cite{mandel1}

The component $\sigma^+$ introduces transitions between
fine structures; $nS_{1/2}\ (m_J=-1/2) \to nP_{1/2}\ (m_J=1/2)$, 
$nS_{1/2}\ (m_J=-1/2) \to nP_{3/2}\ (m_J= 1/2)$, and $nS_{1/2}\ (m_J=
1/2) \to nP_{3/2}\ (m_J=3/2)$. The transitions from $nS_{1/2}$ to
$nP_{3/2}$ are red-detuned, while the transition from $nS_{1/2}$ to
$nP_{1/2}$ is blue-detuned if $\omega_L$ is chosen between the transition
frequencies of $nS_{1/2}\ (m_J=-1/2) \to nP_{3/2}\ (m_J=1/2)$ and
$nS_{1/2}\ (m_J=-1/2) \to nP_{1/2}\ (m_J=1/2)$. Then by adjusting
$\omega_L$ properly, it is possible to cancel the attractive potential
and the repulsive potential associated with these transitions.
The net contribution of the $\sigma^+$ laser beam in this case is an attractive
potential for an atom in the state $nS_{1/2}\ (m_J= 1/2)$. We denote this
potential as $V_+(\bm{x})$ in the following.
Similarly, the $\sigma^-$ component introduces a net attractive potential
$V_-(\bm{x})$, through the transition $nS_{1/2}\ (m_J=-1/2)
\to nP_{3/2}\ (mJ=-3/2)$ on an atom in the state $nS_{1/2}\ (m_J=-1/2)$.

Mandel {\it et al.} \cite{mandel2} took advantage of these state dependent
potentials to implement a two-qubit gate. The potentials acting on $|0
\rangle$ and $|1 \rangle$ are evaluated as
\begin{equation}
\begin{array}{c}
V_{|0 \rangle}(\bm{x}) = \frac{3}{4} V_+(\bm{x}) + \frac{1}{4} V_-(\bm{x})
\vspace{.2cm}\\
V_{|1 \rangle}(\bm{x}) = V_-(\bm{x}).
\end{array}
\end{equation}
By applying the Walsh-Hadamard gate on $|0 \rangle_i |0 \rangle_{i+1}$,
one generates a tensor product state 
$$
\frac{1}{2}(|0 \rangle +|1 \rangle)_i(|0 \rangle + |1 \rangle)_{i+1}.
$$ 
Then by decreasing the phase $\theta$, the state $|0 \rangle$ 
moves with dominating $V_+$ toward $+x$-direction, while $|1 \rangle$ moves
with $V_-$ toward $-x$-direction. Thus it is possible to make $|0 
\rangle_i$ and $|1 \rangle_{i+1}$ collide between the two lattice
points. If they are kept in the common potential well during $t_{\rm hold}$,
the subspace $|0 \rangle_i|1 \rangle_{i+1}$ obtains a dynamical phase
$e^{-i U t_{\rm hold}}$, where $U$ is the on-site repulsive potential
in the potential well. After these two components spent $t_{\rm hold}$
in the potential well, they are brought back to the initial lattice point
by reversing $\theta$, which results in the two-qubit gate operation
\begin{eqnarray}
\lefteqn{|0 \rangle_i |1 \rangle_{i+1}}\nonumber\\
&\to&
\frac{1}{2}(|0 \rangle_i |0 \rangle_{i+1} +e^{-i U t_{\rm hold}}
|0 \rangle_i |1 \rangle_{i+1} \nonumber\\
& &+
|1 \rangle_i |0 \rangle_{i+1} +|1 \rangle_i |1 \rangle_{i+1} ).
\end{eqnarray}
It should be noted, however, that the state dependent potentials act on
all the pairs of the atoms and selective operation of the two-qubit
gate on a particular pair is impossible. Their proposal may be applicable to
generate a highly entangled state for a cluster state quantum computing but
it is not applicable for a circuit model quantum computing.

In the following sections, we propose implementations of a neutral
atom quantum computer, which overcome these difficulties.

\section{near field Fresnel Diffraction Trap}

We show in the present section that atoms are trapped in
an array of NFFD traps, each of which traps a single atom.
An NFFD light is produced if a plane wave is incident to a
screen with an aperture of the radius $a \geq \lambda$.
The ratio $N_F = a/\lambda \geq 1$ is called the Fresnel number.

Let us concentrate on a single neutral atom qubit. 
To make our analysis concrete, we take an alkali metal atom $^{87}$Rb
as an example.
We take qubit basis vectors $|0 \rangle = |F=1, m_F= 1 \rangle$ and 
$|1 \rangle = |F=2, m_F = 1 \rangle$, to be consistent with
single-qubit operations, which involve well-established
two-photon Raman transitions \cite{1q}.

Bandi {\it et al.} \cite{pra} proposed to trap an atom with a microtrap 
employing NFFD light. The trap potential is evaluated by applying the 
Rayleigh-Sommerfeld formula as
\begin{equation}\label{eq:rs}
U(\bm{x}) = -U_0 \frac{|\mathcal{E}(\bm{x})|^2}{E_0^2},
\end{equation} 
where
$$
\mathcal{E}(\bm{x}) = \frac{E_0}{2\pi} \int \int \frac{e^{ikr}}{r} \frac{z}{r}
\left(\frac{1}{r}-ik \right) dx'dy'
$$
and 
$$
U_0 = \frac{3}{8} \frac{\Gamma_e}{|\Delta_{eg}|} \frac{E_0^2}{k^3}.
$$
Here $E_0$ is the amplitude of the incoming plane wave, $k$, its wave
number, the detuning $\Delta_{eg}$ is negative (red-detuned),
and the plane wave is incident to a screen from the $-z$-axis.
The integral is over the aperture region $(x'^2+y'^2 \leq a^2)$.

Figure 1 shows the NFFD trap potential, which is obtained by evaluating
Eq.~(\ref{eq:rs}) numerically. It is found that the distance between the local
minimum of the potential and the aperture
changes from $\sim \lambda$ to $\sim 4 \lambda$ as the aperture radius changes
from $a=\lambda$ to $a=2 \lambda$. Thus the position of the local potential
minimum may be controllable by controlling the aperture radius, which 
may be possible within current technology.
\begin{figure}
\begin{center}
\includegraphics[width=8cm]{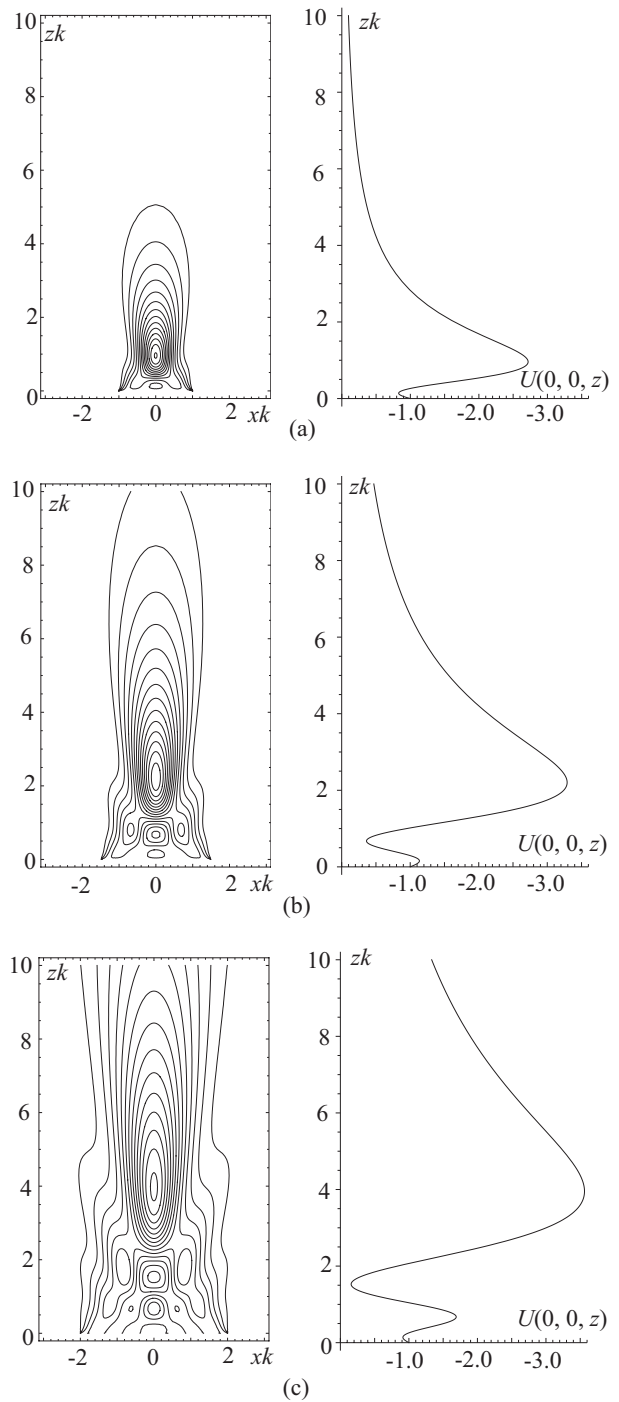}
\end{center}
\caption{Potential profiles of the NEFD trap for (a) $a=\lambda$,
(b) $a=1.5 \lambda$, and (c) $a=2 \lambda$, where $a$ is the aperture
radius and $\lambda$ the wavelength of the incoming laser beam.
The right panel is the contour plot while the left panel shows the
potential profile along the $z$-axis.}
\end{figure}

\begin{figure}
\begin{center}
\includegraphics[width=8.5cm]{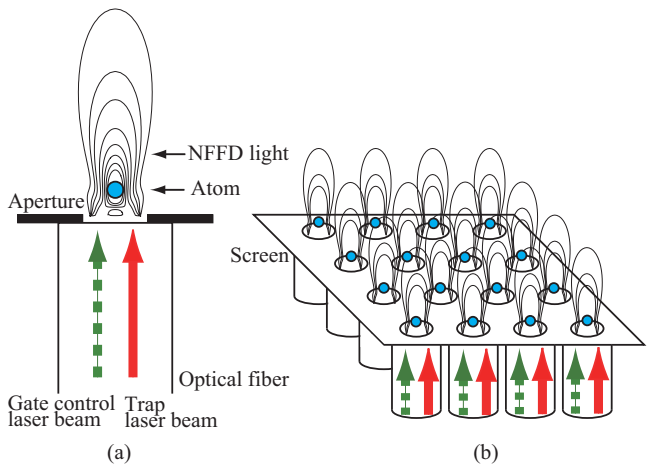}
\end{center}
\label{fig:concept}
\caption{(Color online)
(a) Atom trapped in an NFFD light. Red arrow (solid line) shows
the trap light while the green arrow (broken line) 
shows the laser light required
for 1-qubit gate operations. (b) Example of an array of such traps. 
The trap laser and the
gate control laser are explicitly shown only for the first row.}
\end{figure}
Figure 2 shows an array of optical traps, arranged in a regular 2-d
lattice. It is clear by construction that we have freedom in the
choice of the lattice constant and the lattice shape. We may even
arrange the traps in an irregular form if necessary.
This is a remarkable
new feature, which a conventional optical lattice cannot provide with.
An optical lattice is always regular, whose lattice constant
is on the order of the wavelength of the laser beams.

Attached to each atom is an optical fiber, though
which two laser beams are fed, one is for atom trapping while the
other is to control the hyperfine qubit states of the atom.
It is required to turn on and off the trap laser
individually, which is required to implement a selective two-qubit gate 
as will be shown later.
The gate control laser is also controllable so that 
one-qubit gates may be applied on many qubits simultaneously 
and independently.

\section{Quantum Gate Implementations}

For a quantum system to be a candidate of a general purpose quantum computer,
it is necessary for the system to be able to implement a universal
set of quantum
gates, that is, the set of one qubit gates and almost any two-qubit gate
\cite{Barenco}. Now we explain how these gates are realized in our proposal.

\subsection{One-Qubit Gates}

Let us assume, for definiteness, the trap potential is strong enough so that
inter-atomic interaction is switched off. Along with the NFFD trap light,
there is a gate control laser beam propagating through
the fiber, which is depicted in a green broken line in Fig.~2. 

One-qubit gate is implemented by making use of the two-photon Raman transition
\cite{nakaharaohmi}. Let $E_0$ and $E_1$ be the energy eigenvalues of
the states $|0 \rangle$ and $|1\rangle$, respectively, and
let $E_e$ be the energy eigenvalue of an auxiliary excited state $|e\rangle$
necessary for the Raman transition. Suppose a laser beam with the
frequency $\omega_L$ has been applied to the atom. Let $\hbar
\Delta = \hbar \omega_L-(E_e-E_0)$ be the detuning and
$\Omega_i$ be the Rabi oscillation frequency between the state $|i \rangle
\ (i=0, 1)$ and $|e \rangle$.
Then, under the assumptions $|\Delta| \gg (E_1-E_0)/\hbar, \Omega_i^2/|
\Delta|$, we obtain the effective Hamiltonian
\begin{equation}
H_1 = \frac{1}{2} \epsilon \sigma_z - \frac{\Omega_0 \Omega_1}{4 \Delta}
\sigma_x,
\end{equation}
where 
$$
\epsilon = E_1-E_0 + \frac{\Omega_1^2-\Omega_0^2}{4 \Delta}.
$$
Note that $H_1$ generates all the elements of SU(2) since there are
two $\mathfrak{su}(2)$ generators $\sigma_{x,z}$ in the Hamiltonian and 
their coefficients are controllable.

One-qubit gate implementations with the two-photon Raman transitions have been
already demonstrated \cite{1q}.

\subsection{Two-Qubit Gates}

Although a two-qubit gate operation has been already 
experimentally demonstrated
\cite{mandel2}, a selective gate operation is yet to
be realized. We propose to use a one-dimensional optical lattice or a set of
optical lattices to selectively apply a two-qubit gate in the aforementioned
setting. A one-dimensional optical lattice is made of a pair of 
counterpropagating laser beams with the same linear polarization, in which
the basis states $|0 \rangle=|F=1, m_F=1 \rangle$
and $|1 \rangle=|F=2, m_F=1 \rangle$ are subject to the same trap potential.
Alternatively, hyperfine-state-senstive trap potentials may be introduced
if two counterpropagating laser beams with tilted
polarizations, $\bm{E}_+ \propto e^{ikx} (\hat{\bm{z}} \cos \theta
+ \hat{\bm{y}} \sin \theta)$ and $\bm{E}_- \propto e^{-ikx} 
(\hat{\bm{z}} \cos \theta - \hat{\bm{y}} \sin \theta)$, are superposed.
This results in
\begin{equation}
\bm{E}_+ + \bm{E}_- \propto \sigma^+ \cos (kx-\theta) -
 \sigma^- \cos (kx+\theta),
\end{equation}
where $\sigma^{\pm}$ denote two circular polarizations.
The electric field results in the Stark shifts
\begin{equation}
V_{\pm} (x) \propto \cos^2 (kx\mp \theta)
\end{equation}
and detailed analysis of optical transitions shows that
the effective potentials acting on $|0 \rangle$ and $|1 \rangle$
are
\begin{equation}
\begin{array}{c}
\displaystyle  V_{|0 \rangle}(x) = \frac{1}{4}V_+(x) + \frac{3}{4}V_-(x),
\vspace{.2cm}\\
\displaystyle  V_{|1 \rangle}(x) = \frac{3}{4}V_+(x) + \frac{1}{4}V_-(x),
\end{array}
\end{equation}
respectively,
where use has been made of the following decompositions
\begin{equation}
\begin{array}{c}
\displaystyle |0 \rangle = 
%|F =1, m_F=1 \rangle
 -\frac{1}{2}\left|\frac{3}{2}, \frac{1}{2} \right\rangle  \left|\frac{1}{2}, 
\frac{1}{2} \right\rangle + 
\frac{\sqrt{3}}{2} \left|\frac{3}{2}, \frac{3}{2} \right \rangle
\left| \frac{1}{2}, -\frac{1}{2} \right\rangle,
\vspace{.2cm}\\ 
\displaystyle  |1 \rangle = 
%|F= 2, m_F = 1 \rangle
\frac{\sqrt{3}}{2}\left|\frac{3}{2},\frac{1}{2} \right\rangle
\left|\frac{1}{2}, \frac{1}{2} \right\rangle+
\frac{1}{2}\left|\frac{3}{2}, \frac{3}{2}
\right\rangle
\left|\frac{1}{2}, -\frac{1}{2}\right\rangle.
\end{array}
\end{equation}
It is important to note that $V_{|0 \rangle}(x)$ ($V_{|1
\rangle}(x)$) moves right
(left) if $\theta$ is increased, which implies that the
components $|0 \rangle$ and $|1 \rangle$ move in opposite
directions as $\theta$ is changed.

Suppose a one-dimensional optical lattice, with the potential 
depth considerably
shallower than that of the individual trap potential, is always superposed with
the trap potentials. It is assumed that the lattice constant of the
trap potential array is an integral multiple of the optical lattice period
and the bottom of the trap potential sits in one of the bottoms of
the optical lattice. The atom is in the
motional ground state of the combined potential in an idle time.

Two qubit-gate operation is implemented with six steps.
To simplify our exposition, we assume the trap array is a square
lattice and two atoms, on which the gate acts, are aligned
along one of the primitive lattice vectors. More general cases
will be treated later.
\begin{figure}
\begin{center}
\includegraphics[width=7.5cm]{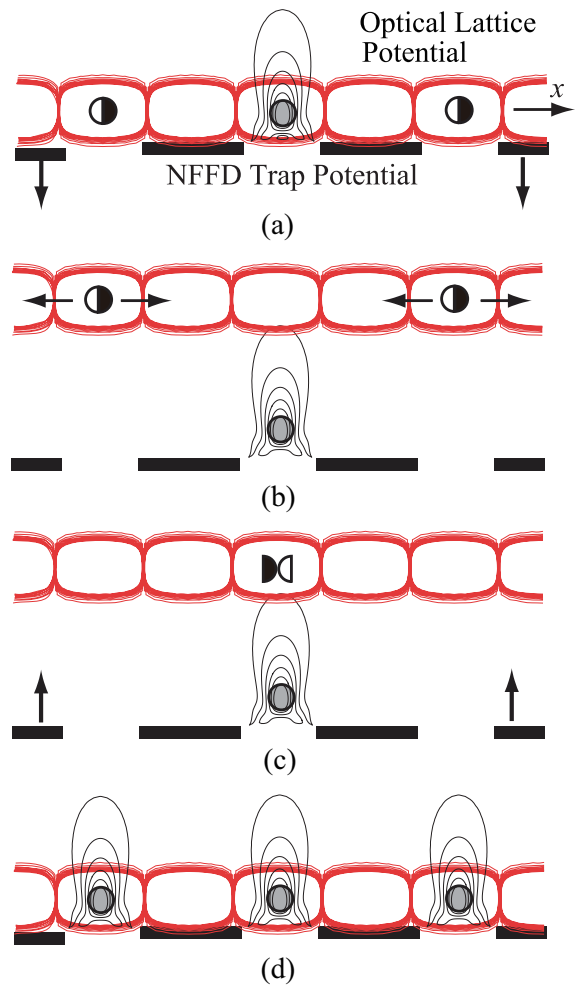}
\end{center}
\caption{
\label{case_B} (Color online)
Schematic diagram of a two-qubit gate operation. 
(a) STEPs 1 and 2. The left and the right atoms are acted by the Walsh-Hadamard
gate in advance so that each of them is in the superposition of
$|0 \rangle$ (white semicircle) and $|1 \rangle$ (black semicircle).
There NFFD trap potentials are turned off adiabatically.
The red contours depict the optical lattice potential along the $x$-axis
while the black coutours show the individual NFFD potential. Bold black
segments show the screen.
(b) STEP 3. The screen supporting the NFFD traps is moved away from the 
optical lattice after traps of the two atoms
are turned off. Atoms that do not participate in the
two-qubit gate operation are withdrawn from the optical lattice.
(c) STEP 4. Two atoms in the superposition of $|0 \rangle$ and $|1 \rangle$
are left in the optical lattice. Now the polarizations of the
counterpropagating laser beams are tilted by $\pm \theta$ so that
the $|0 \rangle$ and $|1 \rangle$ components move in the opposite
directions, which results in the collision of $|0 \rangle$ of one
atom and $|1 \rangle$ of the other. If these components are kept
together for a duration $t_{\rm hold}$, the component obtains
a dynamical phase $e^{-i U t_{\rm hold}}$, where $U$ is the interaction
strength of the two components.
(d) STEPs 5 and 6. 
The screen is put back to the initial position and the NFFD traps of the 
two atoms are turned on subsequently.}
\end{figure}
\begin{enumerate}
\item[STEP 1] %1
The Hadamard gate acts on
each of the two qubits taking part in the gate operation
by manipulating the control laser beams, 
which results in the superpostion of $(|0 \rangle+|1 \rangle)/\sqrt{2}$,

\item[STEP 2] %2
The trap potentials of the two qubits are turned off adiabatically so that 
the resulting motional states remain in the local ground states in the
optical lattice.

\item[STEP 3] %3
Then the trap potential array is shifted
away from the optical lattice. The motional state of a qubit,
which does not take part in the gate operation, ends up with
the ground state of the shifted trap potential provided that the
shift is adiabatically done. Now only the two qubits to be operated
by the gate remain in the optical lattice.

\item[STEP 4]  %4
Polarizations $\theta$ of the counterpropagating laser beams are
controlled so that $|0 \rangle$ of one qubit collides with $|1 \rangle$
of the other qubit. Then the vector 
$|0 \rangle|1 \rangle$ obtains an extra dynamical phase factor
$e^{-iU t_{\rm hold}}$, where $U$ is the energy shift due to the collision 
while $t_{\rm hold}$ is the time during which the two states are in contact
\cite{mandel2}.

\item[STEP 5]  %5
The components $|0 \rangle|1 \rangle$
are put back to their initial positions in the
optical lattice by reversing the change in $\theta$. 

\item[STEP 6]  %6
Then the screen is put back to its
initial position so that the trap potentials overlap again with the
optical lattice. Subsequently, the trap potentials of the two qubits,
which were temporarily switched off, are turned on again. 
One-qubit gates may be applied on the two qubits if necessary.
\end{enumerate}

In case the two atoms are in a general position, not necessarily
along a primitive lattice vector, we may introduce two orthogonal
optical lattices as shown in Fig.~4. 
Then the selected atomic
states $|0 \rangle$ of one atom and $|1 \rangle$ of the other 
meet at the intersection of the two optical lattices
to acquire the extra dynamical phase. It should be noted that 
two-qubit gate operations may be applied simultaneously and independently 
on many pairs of qubits.
\begin{figure}
\begin{center}
\includegraphics[width=7cm]{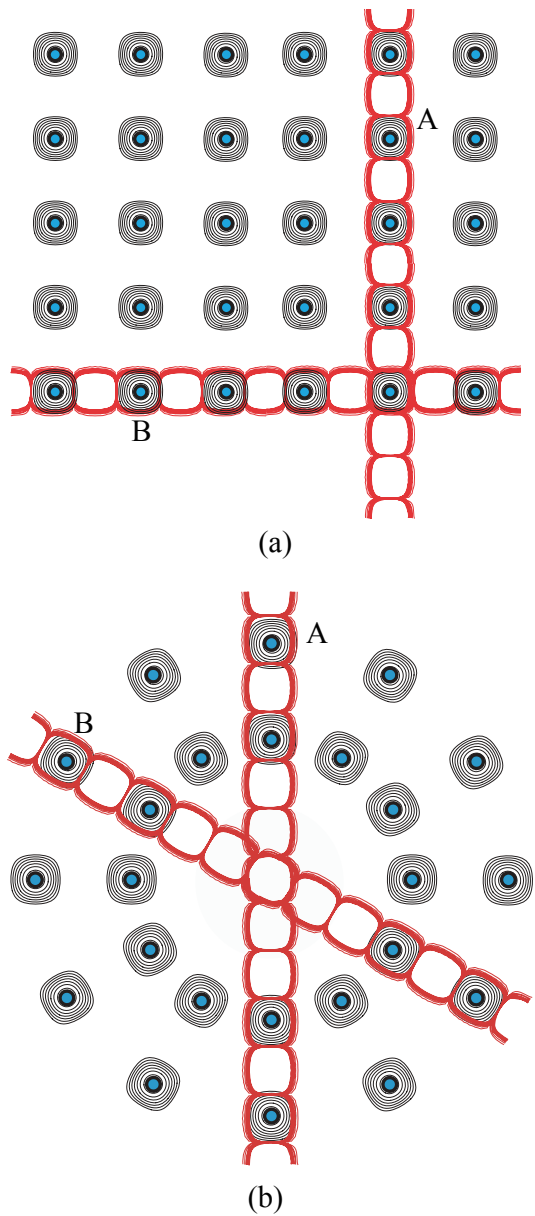}
\end{center}
\caption{(Color online) Two-qubit gate acts on atoms A and B 
in both cases. (a)
Two orthogonal optical lattices (red contours) over two dimensional NFFD traps
(black contours). 
Atoms A and B are in arbitrary positions, not necessarily along one
of the primitive vectors of the array. The screen is looked down from the above.
(b) Radial arrays of NFFD traps. 
The center is the position where interaction
between two qubit states takes place.
\label{2d_optical_lattice}}
\end{figure}

Figure~4~(b) employs a radial
array of NFFD traps. The radial center is the position where interaction
of qubit states takes place. In the radial construction, however,
simultaneous application of two-qubit gates on many pairs of qubits
is difficult.

\section{Variations}

\begin{figure}
\begin{center}
\includegraphics[width=6.5cm]{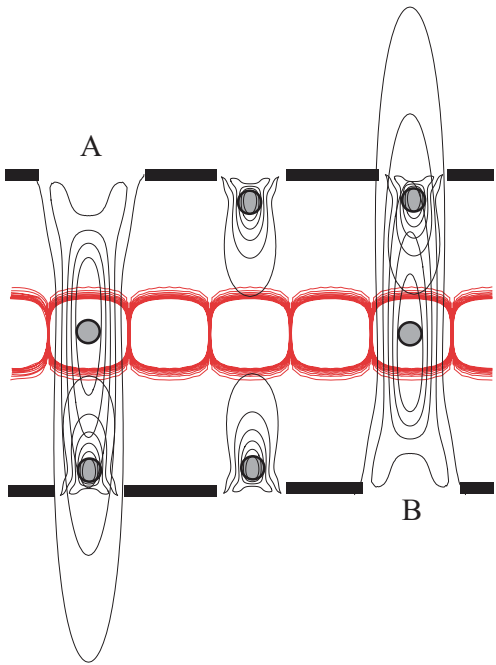}
\end{center}
\caption{
\label{fig:double}
(Color online) Double-layer construction of a neutral atom quantum computer.
Atoms in the initial positions A and B are launched to the optical
lattice (red contours) for entanglement.}
\end{figure}

There are several variations of the current proposal.
Instead of leaving two atoms in the optical lattice and
withdraw the rest of the atoms, we may initialize the
register so that no atoms are present in the optical lattice.
Then a pair of atoms are
sent to the optical lattice by enlarging the size of the
aperture as shown in Fig.~\ref{fig:double}, in which we indicate
a double-layer construction. 
We may double the number of qubits and, at the same time,
reduce the execution time
with this implementation. Note that the vertical position of the atom is
controllable by changing the aperture radius as was demonstrated in
Fig.~1. Then we repeat the two-qubit gate operation (STEPs 4 and 5) outlined
in the previous section to selectively entangle the pair of atoms. 
The size of the aperture can be controlled by employing the well-established
MEMS (MicroElectroMechanical Systems) technology. 
For example, an MEMS shutter for a display is already in a mass production stage \cite{mems1} and it may be employed for physical realization
of our proposal. We note that the
on (off) time of the shutter is 54 (36)~$\mu$s \cite{mems1} and we expect faster
switching time by optimizing its structure for our purpose.~\cite{mems2}

Figure~6 shows the second variation, which involves many red-detuned laser 
beams and pairs of optical tweezers. Atoms are trapped at the intersections
of mutually orthogonal red-detuned laser beams. Each vertical laser
beam has another laser for two-photon Raman transition control employed for 
one-qubit gate operations. Horizontal laser beams are introduced for
vertical confinement. A pair of optical tweezers implements two-qubit
gates. The optical tweezer 1 is $\sigma^+$-polarized to produce a trap 
potential $V_{|0\rangle}(\bm{x})$ for a component $|0 \rangle$, while 
the other (the optical tweezer 2)
is $\sigma^-$-polarized to produce $V_{|1 \rangle}$, which traps
$|1 \rangle$. These optical tweezers are manipulated to pick up
the components $|0 \rangle$ from one atom and $|1 \rangle$ from the
other. These states are put together for a duration $t_{\rm hold}$ as before
to introduce a phase $e^{-i Ut_{\rm hold}}$ in the subspace $|0 \rangle
|1 \rangle$. These components are put back to their initial positions 
after the contact. 
\begin{figure}
\begin{center}
\includegraphics[width=9cm]{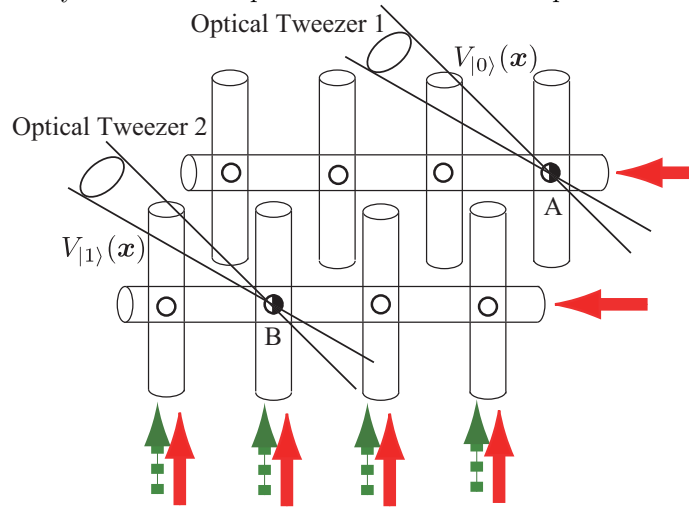}
\end{center}
\caption{(color online)
Lattice of trap potentials made of red-detuned laser beams (red arrows with
solid line). Atoms are
trapped at the intersection of two laser beams. Vertical laser beams
also transmit laser beams (green arrows with broken line)
for two-photon Raman transitions. A pair of optical tweezers 1 and 2
are introduced for two-qubit gate operations.}
\label{fig:rd}
\end{figure}
Similar idea has been proposed in \cite{sanders1,sanders2}, in which 
group-II-like atoms (such as Sr and Yb)
are picked up by a pair of optical tweezers and,
subsequently, they are made to collide with each other. Entanglement
is established by taking advantage of the difference in the $s$-wave
scattering lengths (i.e., interaction strengths) of 
different electronic states of the atom.

Needless to say, we may mix these methods for the best implementation,
which is compatible with the current technology.

\section{Summary and Discussions}

We proposed scalable designs of a neutral atom quantum computer
with an on-demand interaction. A qubit is made of two hyperfine states of
an atom. Associated with each atom, there is an optical fiber, through
which a trap laser and a gate control laser are supplied.
The latter is used to implement one-qubit gates by two-photon
Raman transitions. One-qubit
gate operations are applicable to all the qubits simultaneously and 
independently.
A two-qubit gate may be implemented by leaving two qubit states, $|0 
\rangle$ from one atom and $|1 \rangle$ from the other, in an
optical lattice. By controlling the polarizations of counterpropergating
laser beams, with which the optical lattice is formed, it is possible
to collide these qubit states and introduce an extra dynamical phase in this 
particular two-qubit
state $|0 \rangle |1 \rangle$.
Two-qubit gate operations are also applicable on
many pairs of qubits simultaneously and independently.

We believe our proposal is feasible within the existing technology.
Now we are conducting extensive numerical analysis for each step and
the results will be published elsewhere.

\begin{acknowledgments}
We would like to thank Toshiki Ide for discussion in the
early stage of this work.
MN would like to thank Barry Sanders for drawing his attention to
\cite{sanders1} and \cite{sanders2}.
MN and TO are partly supported by 
Grant-in-Aid for Scientific Research (C) (Grant No.~19540422) from JSPS.
A part of this research is also supported by
``Open Research Center''
Project for Private Universities: Matching fund subsidy from
MEXT (Ministry of Education, Culture, Sports, Science and
Technology).

\end{acknowledgments}


\end{document}